\documentstyle[preprint,prl,aps,epsf,floats]{revtex}
\begin{document}
\draft
\preprint{
\vbox{\halign{&##\hfil         \cr
        & UTPT-98-18           \cr
        & hep-ph/9812505      \cr
        & December 1998        \cr
        & Revised January 1999 \cr}}}

\title{ Associated Production of $\Upsilon$ \\
		and Weak Gauge Bosons in Hadron Colliders } 

\author{Eric Braaten and Jungil Lee}
\address{Physics Department, Ohio State University, Columbus OH 43210, USA}

\author{Sean Fleming}
\address{Physics Department, University of Toronto, 
	Toronto, Ontario M5S 1A7, Canada}

\maketitle
\begin{abstract}
We calculate the rate of production of $W^\pm + \Upsilon$ 
and $Z^0 + \Upsilon$ in hadron colliders.
We find the cross sections  
for $W^\pm + \Upsilon$ and $Z^0 +\Upsilon$ to be roughly 
0.45 pb and 0.15 pb at the Tevatron and roughly 
4 pb and 2 pb at the LHC.
The dominant production mechanism involves the binding  of a 
color-octet $b \bar b$ pair into a P-wave bottomonium state 
which subsequently decays into $\Upsilon$.
The purely leptonic decay modes of $\Upsilon$, $W^\pm$, and $Z^0$
provide signatures with small backgrounds.
These events may be observable in Run II at the Tevatron, 
and they should certainly be observable at the LHC.
\end{abstract}

\vfill \eject

\narrowtext

In high energy collider experiments, the particles with the cleanest
signatures are those with purely leptonic decay modes.  
They include the charged weak gauge bosons $W^\pm$, 
which decay via $W \to \ell \nu$ where $\ell$ is an electron or muon, 
and the neutral weak gauge boson $Z^0$,
which decays via $Z^0 \to \ell^+ \ell^-$.  
They also include the $J^{PC} = 1^{--}$ quarkonium states, 
such as the charmonium state $J/\psi$ and the 
bottomonium state $\Upsilon$, which decay into $\ell^+ \ell^-$.
Purely leptonic decay modes are particularly useful in hadron colliders, 
because they provide an enormous suppression of the background.
Events involving particles with such decay modes  
can be employed both as a probe of the production process
and as a lamppost under which to search for new physics.
In this letter, we calculate the rate 
in the Standard Model
for the associated production
of the bottomonium state $\Upsilon$ with $W^\pm$ or $Z^0$ 
at the Tevatron proton-antiproton collider
and at the LHC proton-proton collider.  
We find that  
the dominant production mechanism for both $W^\pm + \Upsilon$
and $Z^0 + \Upsilon$ involves the binding of a color-octet
$b \bar b$ pair into a P-wave bottomonium state 
which subsequently decays into $\Upsilon$.
The purely leptonic decays of the $\Upsilon$, $W^\pm$, and $Z^0$
provide clean signatures for these events.
The cross sections for $W^\pm + \Upsilon$ 
and $Z^0 + \Upsilon$ may be too small for these events
to be observed at the Tevatron, 
but they should certainly be seen at the LHC.

The associated production of $W^\pm$ or $Z^0$ and
a $J/\psi$ has been studied previously, but only for $J/\psi$'s
with large transverse momentum, where the process is dominated
by gluon fragmentation into $J/\psi$ \cite{B-F-R}. 
The differential cross sections for $W^\pm + J/\psi$ 
and $Z^0 + J/\psi$ peak at small transverse momentum of the $J/\psi$.
Unfortunately, if the $J/\psi$ is produced with
small transverse momentum, one or more of the leptons from its decay 
is likely to have insufficient energy to be identified.  
Because the mass of the $\Upsilon$ is larger 
than that of the $J/\psi$ by a factor of 3,
the cross sections for $W^\pm + \Upsilon$ 
and $Z^0 + \Upsilon$ are much smaller 
than those for $W^\pm + J/\psi$ and $Z^0 + J/\psi$. 
However the larger mass of the $\Upsilon$ 
makes it possible to observe the leptons from its decay, even if 
the $\Upsilon$ is produced at small transverse momentum.
It is therefore possible to measure the total cross sections for 
$W^\pm + \Upsilon$ and $Z^0 + \Upsilon$.

The production of $\Upsilon$ requires the creation of a $b \bar b$ pair
at short distances of order $1/m_b$ or smaller.  The $b \bar b$ pair
subsequently forms a color-singlet bound state, which can either be 
$\Upsilon$ or a higher bottomonium state that decays 
into $\Upsilon$.
The probability of binding depends on the bottomonium state and 
on the color and angular momentum quantum numbers of the $b \bar b$ pair.
In the NRQCD factorization approach to 
inclusive quarkonium production \cite{B-B-L}, 
these probabilities are parameterized by 
nonperturbative matrix elements in nonrelativistic QCD (NRQCD). 
The matrix elements scale in a definite way 
with the typical relative velocity $v$ of the bottom quark in bottomonium.
Its value is roughly $v^2 \approx 1/10$. 
The NRQCD scaling rules can be used to estimate the order of magnitude
of the binding probability, with every factor of $v^2$
corresponding to suppression by an order of magnitude. 

We first consider the direct production of $\Upsilon$.
The probability of forming an $\Upsilon$ is largest 
for the $b \bar b_1(^3S_1)$ state, which is a color-singlet
configuration with angular-momentum quantum numbers $^3S_1$.  
It can be parameterized by the NRQCD matrix element 
$\langle {\cal O}^\Upsilon_1(^3S_1) \rangle$, which is
proportional to the square of the wavefunction at the origin and can be
extracted phenomenologically from the leptonic width of the $\Upsilon$:
$\langle {\cal O}^\Upsilon_1(^3S_1) \rangle$ = 7  GeV$^3$.
According to the velocity-scaling rules of NRQCD,
the $b \bar b$ configurations with
the next largest probabilities for binding into an $\Upsilon$  
are the color-octet states 
$b \bar b_8(^3S_1)$, $b \bar b_8(^1S_0)$, 
and $b \bar b_8(^3P_J)$  \cite{B-B-L}.
Their binding probabilities are suppressed by $v^4$ relative to
$b \bar b_1(^3S_1)$.
If the cross section for color-octet $b \bar b$ pairs was 
comparable to that for $b \bar b_1(^3S_1)$,
then the color-octet contributions to the cross section for direct 
$\Upsilon$ production would be suppressed by two orders of magnitudes.
However in hadron colliders, the cross section for color-octet 
$b \bar b$ pairs is much larger than that for $b \bar b_1(^3S_1)$,
and this can compensate for the smaller binding probability.
The cross section for $b \bar b_8(^3S_1)$ is particularly large, 
because it can be produced by the decay of an off-shell gluon.
There are as yet no reliable phenomenological determinations of the 
matrix elements $\langle {\cal O}^\Upsilon_8(^3S_1) \rangle$,
$\langle {\cal O}^\Upsilon_8(^1S_0) \rangle$, and 
$\langle {\cal O}^\Upsilon_8(^3P_0) \rangle$.
In their analysis of $\Upsilon$ production at the Tevatron 
\cite{Cho-Leibovich}, Cho and Leibovich used the value
$\langle {\cal O}^\Upsilon_8(^3S_1) \rangle$ 
$\approx$ 0.006 GeV$^3$, which they obtained by taking a 
phenomenological value for 
$\langle {\cal O}^{J/\psi}_8(^3S_1) \rangle$ with large error bars
and scaling it by $(m_b v_b)^3/(m_c v_c)^3$.
They extracted a value for the linear combination
$\langle {\cal O}^\Upsilon_8(^1S_0) \rangle/5$ + 
$\langle {\cal O}^\Upsilon_8(^3P_0) \rangle/m_b^2$ 
by fitting CDF data for bottomonium production \cite{CDF-Upsilon}.
They obtained a central value 0.008 GeV$^3$ with a statistical uncertainty
that was greater than 100\%. 
Thus the uncertainties in the color-octet matrix elements are 
probably at least an order of magnitude.

We next consider the indirect production of $\Upsilon$ by the 
binding of the $b \bar b$ pair into a higher bottomonium state
that subsequently decays into $\Upsilon$.  The higher states
include $\Upsilon(2S)$ and $\Upsilon(3S)$, which decay into 
$\Upsilon$ with branching fractions of about 31\% and 13\%, 
respectively.  Since these states have the same quantum numbers 
$J^{PC} = 1^{--}$ as $\Upsilon$, 
their NRQCD matrix elements scale in the same way with $v$
and will therefore be comparable in magnitude to those for $\Upsilon$.
For example, the binding probabilities of $b \bar b_1(^3S_1)$
into $\Upsilon(2S)$ and $\Upsilon(3S)$ are both about 
40\% smaller than for $\Upsilon$.  
The effect of the feeddown from higher bottomonium states on the 
production of $\Upsilon$ from $b \bar b_1(^3S_1)$ can be
taken into account in the NRQCD factorization formulas by replacing
$\langle {\cal O}^\Upsilon_1(^3S_1) \rangle$ by
\begin{equation}
\sum_H B(H \to \Upsilon X) \langle {\cal O}^H_1(^3S_1) \rangle 
\;=\; 8 \; {\rm GeV}^3,
\label{O-sing:inc}
\end{equation}
where we have added the contributions from $\Upsilon$ (with $B=1$),
$\Upsilon(2S)$, and $\Upsilon(3S)$.
Including the feeddown from $\Upsilon(2S)$ and $\Upsilon(3S)$ 
therefore increases the cross section by about 20\%. 

Bottomonium states with different $J^{PC}$ quantum numbers 
can have a more dramatic effect on the cross section, because they
receive contributions from parton processes with different selection rules.
The only process whose binding probability is of the same order in $v$ as
$b \bar b_1(^3S_1)$ into $\Upsilon(nS)$
is $b \bar b_1(^1S_0)$ into $\eta_b(nS)$.
By the heavy-quark spin symmetry of NRQCD, the matrix element
$\langle {\cal O}^{\eta_b(nS)}_1(^3S_1) \rangle$  differs from that of 
$\langle {\cal O}^{\Upsilon(nS)}_1(^1S_0) \rangle$ 
only by a spin factor of 1/3.  Including the contributions from 
$\eta_b(2S)$ and $\eta_b(3S)$ and assuming that
$B(\eta_b(nS) \to \Upsilon X)$ is comparable to the branching fraction for 
hadronic transitions from $\Upsilon(nS)$ to $\Upsilon$,
we obtain the estimate
$\sum_H B(H \to \Upsilon X) \langle {\cal O}^H_1(^1S_0) \rangle$ 
$\approx$ 0.3 GeV$^3$.

The binding probabilities for P-wave bottomonium states
are suppressed by $v^2$ relative to $\Upsilon$.
For $\chi_{bJ}(nP)$, the $b \bar b$ configurations 
with binding probabilities suppressed by only $v^2$ are
$b \bar b_1(^3P_J)$ and $b \bar b_8(^3S_1)$.
By heavy-quark spin symmetry, the color-singlet matrix elements
$\langle {\cal O}^{\chi_{bJ}(nP)}_1(^3P_J) \rangle$
differ for $J=0,1,2$ only by a spin factor of $2J+1$.
Their values can be determined from the wavefunction 
in nonrelativistic potential models.  
We adopt the values used in the analysis of Ref. \cite{Cho-Leibovich}.  
Weighting the matrix elements for $\chi_{bJ}(1P)$  and $\chi_{bJ}(2P)$
by their measured branching fractions into $\Upsilon$, we obtain
$\sum_H B(H \to \Upsilon X) \langle {\cal O}^H_1(^3P_J) \rangle/m_b^2$ 
$\approx$ (0.003,0.2,0.2) GeV$^3$ for $J=(0,1,2)$.  We have included 
a factor of $1/m_b^2$ in the P-wave matrix elements so that they
have the same dimensions as S-wave matrix elements.
The relative magnitudes of the matrix elements can then be
interpreted as relative binding probabilities.
The color-octet matrix elements 
$\langle {\cal O}^{\chi_{bJ}(nP)}_8(^3S_1) \rangle$
differ for $J=0,1,2$ only by a spin factor of $2J+1$.  In the analysis  
of $\Upsilon$ production at the Tevatron by Cho and Leibovich 
\cite{Cho-Leibovich}, these matrix elements were determined by 
fitting the CDF data on bottomonium production \cite{CDF-Upsilon}.
Weighting the matrix elements by their measured
branching fractions into $\Upsilon$, we obtain
\begin{equation}
\sum_H B(H \to \Upsilon X) \langle {\cal O}^H_8(^3S_1) \rangle 
\;\approx\; 0.4 \; {\rm GeV}^3.
\label{O-oct:inc}
\end{equation}
We have included the contributions from $\chi_{bJ}(1P)$ with $J=1,2$ and
from $\chi_{bJ}(2P)$ with $J=0,1,2$.  The $1P$ states 
account for about 70\% of the total.  The value (\ref{O-oct:inc})
is more than an order of magnitude larger than the estimate 
by Cho and Leibovich of the matrix element
$\langle {\cal O}^\Upsilon_8(^3S_1) \rangle$ 
that contributes to direct $\Upsilon$ production.
The uncertainty in (\ref{O-oct:inc}) is probably at least a factor of 3, 
because the analysis of $\Upsilon$ production has
substantial theoretical uncertainties.
The weighted matrix element in (\ref{O-oct:inc}) is smaller
than the corresponding color-singlet term in (\ref{O-sing:inc})
by an order of magnitude.
This smaller binding probability is more than 
compensated by the much larger cross section for $b \bar b_8(^3S_1)$ 
in hadron colliders. 
Thus the binding of $b \bar b_8(^3S_1)$ 
into $\chi_{bJ}$ followed by its decay into $\Upsilon$ 
gives a large contribution to the total $\Upsilon$ cross section.

We now consider 
the production of $W^\pm + \Upsilon$  in a hadron collider.  This process
proceeds through parton collisions that produce 
final states including $W + b \bar b$.  The relative importance of 
the various contributions depends on the magnitude of the 
NRQCD matrix elements and on the size of the parton cross sections.
The only parton cross sections that can be
calculated relatively easily are those that can be taken into account
via $2 \to 3$ parton processes of the form $i j \to W + b \bar b$.
Fortunately, they account for the largest contributions. 
The $2 \to 3$ parton processes that produce $W^+ + b \bar b$ are
\begin{itemize}
\item W1:
$u \bar d, c \bar s \to W^+ + b \bar b_8(^3S_1)$ 
via the decay of a virtual gluon into $b \bar b$,

\item W2:
$c \bar b \to W^+ + b \bar b_1(^3S_1,^1S_0,^3P_J)$ 
involving the exchange of a virtual gluon,

\item W3:
$u \bar d, c \bar s \to W^+ + b \bar b_1(^3S_1)$ 
via the decay of a virtual photon into $b \bar b$.

\end{itemize}
We have omitted processes that are suppressed by a tiny Kobayashi-Maskawa
factor. Process W1 is the only color-octet contribution with a cross section 
of order $\alpha_s^2 \alpha_w$.   All other color-octet contributions 
are suppressed by an additional factor of $\alpha_s$. 
Color-singlet contributions involving QCD interactions 
are suppressed by an additional factor of $\alpha_s^2$.  
One such process, 
$g g \to W^+ + b \bar b_1(^3S_1) + b \bar c$, which has a cross section 
of order $\alpha_s^4 \alpha_w$, is dominated by the splitting 
of the colliding gluons into a collinear $b \bar b$ pair 
and a collinear $c \bar c$ pair.  It can be taken into account 
through process W2.  We expect this 
process to provide a reasonable estimate of the 
color-singlet contributions involving QCD interactions, 
because it is the only such process that is enhanced by a factor of
$\log(M_W/m_b) \log(M_W/m_c)$.
Process W3 is a color-singlet contribution that involves only
electroweak  interactions and has a cross section 
of order $\alpha^2 \alpha_w$. 

We proceed to calculate the cross section for production of 
$W^\pm + \Upsilon$ at the Tevatron.
We use the leading order GRV-94 parton distributions, 
and set the factorization and renormalization scales equal. 
We neglect the masses of the $u$, $d$, $s$, and $c$ quarks,
and we use the value 4.7 GeV for the $b$ quark mass.
The cross section in $p \bar p$ collisions at center-of-mass energy
1.8 TeV is 
\begin{equation}
\sum_\pm \sigma(W^\pm + \Upsilon) \;=\;
1.6 \; {\rm fb}\; 
{\sum B \langle {\cal O}_1(^3S_1) \rangle \over 8 \; {\rm GeV}^3}
\;+\; 450 \; {\rm fb} \; 
{\sum B \langle {\cal O}_8(^3S_1) \rangle \over 0.4 \; {\rm GeV}^3}.
\label{sig-W}
\end{equation}
We have omitted the contributions from the color-singlet $^1S_0$
and $^3P_J$ matrix elements, because they are several orders of 
magnitude smaller than the color-singlet $^3S_1$ term.
The color-singlet $^3S_1$ term is completely dominated by the 
purely electroweak process W3.  However it is overwhelmed by
the color-octet $^3S_1$ contribution from process W1.
While the color-singlet matrix element is accurately normalized by the 
leptonic decay width of the $\Upsilon$, there is a large uncertainty 
in the normalization of the color-octet matrix element.
The theoretical error due to higher order perturbative corrections 
can be estimated by varying the factorization and renormalization scales 
by a factor of 2 and is roughly 35\%.
The theoretical error due to the uncertainty in the $b$ quark mass
can be estimated by varying $m_b$ by 0.25 GeV
and is roughly 25\%.

To obtain the cross section in the purely leptonic decay channels,
we multiply by the branching fractions 5\% for 
$\Upsilon \to \ell^+ \ell^-$ and 22\% for  $W \to \ell \nu$.
Given the integrated luminosity of 110 pb$^{-1}$ in Run I of the Tevatron,
there should be about 0.5 
events in the leptonic decay channels.
To estimate the number of events that would actually be detected,
we must multiply by the detector acceptances and efficiencies.
The product of the acceptances and efficiencies
for the CDF detector was about 26\% for $W \to e \nu$ \cite{CDF-WZ}
and about 12\% for $\Upsilon \to \mu^+ \mu^-$ \cite{CDF-Upsilon}.
Thus after allowing for acceptances and efficiencies, 
the cross section is about two orders of magnitude too small to 
be observable in the data from Run I. 

We next consider the production of $Z^0 + \Upsilon$ in a hadron collider. 
The $2 \to 3$ parton processes that produce $Z^0 + b \bar b$ are
\begin{itemize}
\item Z1:
$u \bar u, d \bar d, s \bar s, c \bar c \to Z^0 + b \bar
b_8(^3S_1,^1S_0,^3P_J)$, 

\item Z2:
$g g \to Z^0 + b \bar b_8(^3S_1,^1S_0,^3P_J)$,

\item Z3:
$g g \to Z^0 + b \bar b_1(^3S_1,^1S_0,^3P_J)$,

\item Z4:
$b \bar b \to Z^0 + b \bar b_1(^3S_1,^1S_0,^3P_J)$,

\item Z5:
$u \bar u, d \bar d, s \bar s, c \bar c \to Z^0 + b \bar b_1(^3S_1)$ 
via the decay of a virtual photon into $b \bar b$.

\end{itemize}
Processes Z1 and Z2 are color-octet contributions with 
cross sections of order $\alpha_s^2 \alpha_w$.  
Process Z3 is the only color-singlet contribution with 
a cross section of order $\alpha_s^2 \alpha_w$. 
Another color-singlet contribution  
$g g \to Z^0 + b \bar b_1(^3S_1) + b \bar b$, 
which has a cross section of order 
$\alpha_s^4 \alpha_w$, can be taken into account through process Z4.  
Finally, process Z5 is an electroweak color-singlet contribution 
with a cross section of order $\alpha^2 \alpha_w$.

The cross section for production of 
$Z^0 + \Upsilon$ in $p \bar p$ collisions at the Tevatron with
center-of-mass energy 1.8 TeV is 
\begin{equation}
\sigma(Z^0 + \Upsilon) \;=\;
8.5 \; {\rm fb}  \;
{\sum B \langle {\cal O}_1(^3S_1) \rangle \over 8 \; {\rm GeV}^3}
\;+\; 150 \; {\rm fb} \;  
{\sum B \langle {\cal O}_8(^3S_1) \rangle \over 0.4 \; {\rm GeV}^3}.
\label{sig-Z}
\end{equation}
We have calculated all contributions to direct $\Upsilon$ production
with binding probabilities of order $v^4$ or larger
and all indirect contributions involving binding probabilities 
of order $v^2$ or larger.  
All those terms that are not shown explicitly in (\ref{sig-Z})
give contributions that are more than an order of magnitude smaller than the 
color-singlet $^3S_1$ term.
The color-singlet $^3S_1$ contribution is dominated by processes Z3 and Z5,
which contribute about 85\% and 15\%, respectively. 
However it is overwhelmed by
the color-octet $^3S_1$ term, which is dominated by process Z1. 
While the color-singlet matrix element is accurately normalized, 
there is a large uncertainty 
in the normalization of the color-octet matrix element.
We estimate the theoretical errors 
from higher order perturbative corrections 
and from the uncertainty in the $b$ quark mass to be roughly 35\%
and 25\%, respectively.

To obtain the cross section in the purely leptonic decay channels,
we must multiply by the branching fractions 5\% for 
$\Upsilon \to \ell^+ \ell^-$ and 7\% for $Z^0 \to \ell^+ \ell^-$.
Given the integrated luminosity of 110 pb$^{-1}$ in Run I of the Tevatron,
there should be about 0.06
events in the leptonic decay channels.
The product of the acceptances and efficiencies
of the CDF detector was about 30\% for $Z^0 \to e^+ e^-$ \cite{CDF-WZ}
and about 12\% for $\Upsilon \to \mu^+ \mu^-$  \cite{CDF-Upsilon}.
Thus after allowing for acceptances and efficiencies, 
the cross section is several orders of magnitude too small to 
be observable in the data from Run I. 

In Run II of the Tevatron,  
the increase of the center-of-mass energy to 2.0 TeV
will increase the cross sections for $W^\pm + \Upsilon$ 
and $Z^0 + \Upsilon$ by about 10\%.
With an integrated luminosity of 2000 pb$^{-1}$,
the number of events in the purely leptonic decay channels
should be about 10 for $W^\pm + \Upsilon$
and about 2 for $Z^0 + \Upsilon$.  
The upgrades of the CDF and D0 detectors
should increase the acceptances and efficiencies for observing these
events. 
Given that the uncertainties in the cross sections 
are at least a factor of 3, there is a possibility 
that these events could be observed in Run II.
An observation of these events at a much larger rate than predicted 
could be evidence for a heavy particle that has a substantial
branching fraction into $W^\pm + b \bar b$ or $Z^0 + b \bar b$. 

\begin{figure}[htbp]
\epsfxsize=12.5cm
\hfil\epsfbox{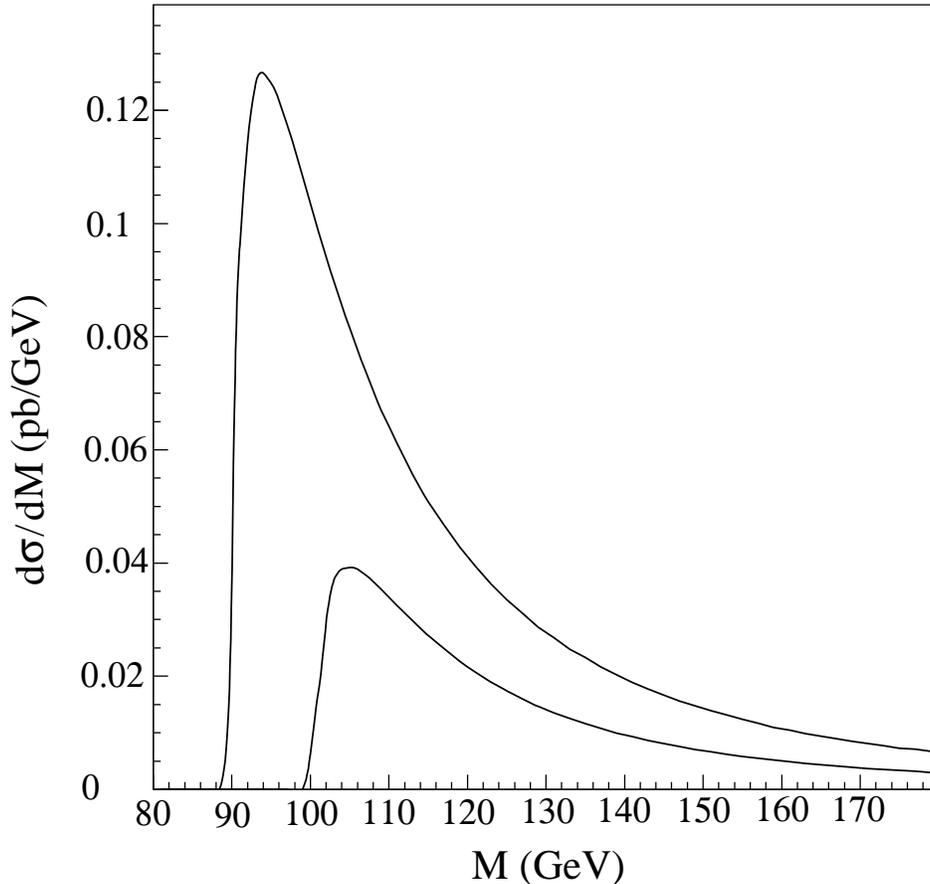}\hfill 
\caption{ The invariant mass distributions 
$d \sigma/dM_{W \Upsilon}$ (upper curve)
and $d \sigma/dM_{Z \Upsilon}$  (lower curve)
for the production of $W^\pm + \Upsilon$ and $Z^0 + \Upsilon$ 
in $p p$ collisions at center-of-mass energy 14~TeV.
} 
\label{wzfig}
\end{figure}

The cross sections for production of $W^\pm + \Upsilon$ and
$Z^0 + \Upsilon$ in $p p$ collisions at the LHC with
center-of-mass energy 14 TeV is 
\begin{eqnarray}
\sum_\pm \sigma(W^\pm + \Upsilon) &=&
10 \; {\rm fb}\; 
{\sum B \langle {\cal O}_1(^3S_1) \rangle \over 8 \; {\rm GeV}^3}
\;+\; 4000 \; {\rm fb} \; 
{\sum B \langle {\cal O}_8(^3S_1) \rangle \over 0.4 \; {\rm GeV}^3},
\label{sig-W:LHC}
\\
\sigma(Z^0 + \Upsilon) &=&
500 \; {\rm fb}  \;
{\sum B \langle {\cal O}_1(^3S_1) \rangle \over 8 \; {\rm GeV}^3}
\;+\; 1300 \; {\rm fb} \;  
{\sum B \langle {\cal O}_8(^3S_1) \rangle \over 0.4 \; {\rm GeV}^3}.
\label{sig-Z:LHC}
\end{eqnarray}
With an integrated luminosity of 10 fb$^{-1}$,
the number of events in the purely leptonic decay channels
should be about 440 for $W^\pm + \Upsilon$
and about 70 for $Z^0 + \Upsilon$. 
Even after allowing for detector acceptances and efficiencies,
there should be enough events to make these processes observable.
In Figure~\ref{wzfig}, we plot the invariant mass distributions 
$d \sigma/dM_{W \Upsilon}$ (summed over $W^\pm$) and 
$d \sigma/dM_{Z \Upsilon}$.  
They peak at only a few GeV above the thresholds
of 89.8 GeV for $W^\pm + \Upsilon$ and 100.6 GeV for $Z^0 + \Upsilon$.

The cross sections for $W^\pm + \Upsilon$ and $Z^0 + \Upsilon$
are both dominated by the $\langle {\cal O}_8(^3S_1) \rangle$ term.
The uncertainty in the value of $\sum B \langle {\cal O}_8(^3S_1) \rangle$
leads to a substantial uncertainty in the normalizations of the 
cross sections.  However this uncertainty cancels in the ratio 
$\sum_\pm \sigma(W^\pm + \Upsilon)/\sigma(Z^0 + \Upsilon)$,
which is predicted to be about 3
at the Tevatron and about 2 at the LHC.  
If a significant deviation from this 
prediction is observed at the LHC, 
it might be evidence for an additional contribution 
from a heavy particle that decays into 
$W^\pm + b \bar b$ or $Z^0 + b \bar b$.

The production of $W^\pm + \Upsilon$ and $Z^0 + \Upsilon$ provide
lampposts under which one can look for new physics.
The most promising possibility is to search for a charged Higgs 
via the decay $H^+ \to W^+ + \Upsilon$.
The decay rate of the charged Higgs into $W + b \bar b$
is enhanced by the Yukawa coupling of the 
Higgs to a virtual top quark.
If the mass of the charged Higgs is in the range between
140 GeV and $t \bar b$ threshold
and if the Higgs mixing parameter $\tan \beta$ is small, then
$W + b \bar b$ may be the largest single decay mode \cite{M-R-W}.
The decay rate of the charged Higgs into $W + \Upsilon$
was first calculated by Grifols, Gunion, and Mendez \cite{G-G-M}.  
For small $\tan \beta$,
the branching fraction $B(H^+ \to W^+ + \Upsilon)$ 
ranges from about $10^{-4}$ if the Higgs mass is just above the
$W^+ + \Upsilon$ threshold to about $10^{-3}$ if the Higgs mass is 
just below the $t \bar b$ threshold.
 
In a hadron collider, most of the standard production mechanisms 
for a charged Higgs in the mass range below $t \bar b$ threshold
involve the production of an additional very massive particle \cite{HHG}.
The standard production mechanisms for $H^+$
include $t \bar t$ production followed by the decay $t \to H^+ b$,
$\bar t H^+$ production, $W^- H^+$ production, and $H^- H^+$ production.
Because of the additional very massive particle, 
events in which a charged Higgs decays into $W + \Upsilon$ will 
be easily distinguished 
from $W + \Upsilon$ events produced by Standard Model processes.
There is one potentially significant production process for a charged Higgs
that could result in events that resemble
Standard Model $W + \Upsilon$ events.  That process is
$q b \to q' b H^+$, which proceeds through a Feynman diagram
that involves a virtual $W$ and a virtual top quark \cite{M-O}. 

We conclude that it should be possible to observe 
the production of $W^\pm + \Upsilon$ and $Z^0 + \Upsilon$ 
from Standard Model proceses at the LHC.  The clean experimental 
signatures for these events also makes them valuable lamposts
under which to search for heavy particles that decay into
$W^\pm + b \bar b$ or $Z^0 + b \bar b$.

This work was supported in part by the U.S.
Department of Energy Division of High Energy Physics under
Grant DE-FG02-91-ER40690, by KOSEF, and by NSERC.  
E.B. would like to thank Kwan Lai for valuable discussions.
S.F. would like to thank the Caltech theory group for 
their hospitality while some of this work was being carried out.  
We thank M. Mangano for pointing out a numerical 
error in a previous version of this paper.


%

\end{document}